\documentclass[aps,twocolumn,showpacs,preprintnumbers,amsmath,amssymb,showkeys,floatfix,superscriptaddress]{revtex4}

\usepackage{graphicx,epsf,epsfig}
\usepackage{dcolumn}
\usepackage{bm}
\usepackage{color}
\usepackage{ulem}

\begin{document}


\title{Investigation of Gd$_{3}$N@C$_{2n}$  (40 $\leq$ $n$ $\leq$ 44) family by Raman and inelastic electron tunneling spectroscopy}
\author{Brian G. Burke}
\author{Jack Chan}
\author{Keith A. Williams}
\email[Corresponding Author: ]{kwilliams@virginia.edu}
\affiliation{Department of Physics, University of Virginia, Charlottesville, Virginia 22904}
\author{Jiechao Ge}
\author{Chunying Shu}
\author{Wujun Fu}
\author{Harry C. Dorn}
\affiliation{Department of Chemistry, Virginia Polytechnic Institute and State University, Blacksburg, Virginia 24061}
\author{James G. Kushmerick}
\affiliation{National Institute of Standards and Technology, Gaithersburg, Maryland 20899}
\author{Alexander A. Puretzky}
\author{David B. Geohegan}
\affiliation{Materials Science and Technology Division, Center for Nanophase Materials Sciences, Oak Ridge National Laboratory, Oak Ridge, Tennessee 37831}

\begin{abstract}
The structure and vibrational spectrum of Gd$_{3}$N@C$_{80}$ is studied through Raman and inelastic electron tunneling spectroscopy (IETS) as well as density functional theory (DFT) and universal force field (UFF) calculations. Hindered rotations, shown by both theory and experiment, indicate the formation of a Gd$_{3}$N$-$C$_{80}$ bond which reduces the ideal icosahedral symmetry of the C$_{80}$ cage. The vibrational modes involving the movement of the encapsulated species are a fingerprint of the interaction between the fullerene cage and the core complex. We present Raman data for the Gd$_{3}$N@C$_{2n}$ (40 $\leq$ $n$ $\leq$ 44) family as well as Y$_{3}$N@C$_{80}$, Lu$_{3}$N@C$_{80}$, and Y$_{3}$N@C$_{88}$ for comparison. Conductance measurements have been performed on Gd$_{3}$N@C$_{80}$ and reveal a Kondo effect similar to that observed in C$_{60}$.
\end{abstract}

\pacs{Valid PACS appear here}
\maketitle

\section{Introduction}
Endohedral fullerenes (endofullerenes), due to their unique electronic and geometric structure, have been attracting the interest of chemists, physicists, and material scientists since they were first synthesized in 1985 \cite{heath}. Fullerenes can act as hosts that encapsulate other atoms, molecules, or atomic clusters. The resulting endofullerenes have initiated considerable research because the entrapped atoms bring with them an array of useful physical properties. The potential applications include electronic devices \cite{kobayashi}, organic solar cells \cite{holloway}, spin-based quantum computing \cite{harneit,larsson}, and medicine \cite{kato,yang,zuo,zhang,wilson,wilson2,mikawa,iezzi,okumura,bolskar,fatouros,zhang2,sitharaman,chaur,shu,shu2}, specifically contrast agents for magnetic resonance imaging (MRI) as well as radiopharmaceutical candidates \cite{yang,zuo}. Endofullerenes created by an arc-vaporization technique \cite{stevenson} have been known to encapsulate up to three individual metal atoms as well as clusters of the form M$_{3}$N, M$_{3}$C$_{2}$, M$_{2}$C$_{2}$, and M$_{4}$O$_{2}$ (M = Sc, Y, and several lanthanides) \cite{yang}. The possible isolation of magnetic or radioactive molecules from the surrounding environment have made endofullerenes interesting for many fields. In particular, the Gd$_{3}$N@C$_{2n}$ family has received much interest due to their application as MRI contrast agents and the confinement of three paramagnetic Gd ions in a relatively small space ($\sim$ 1 nm).

The vibrational modes involving the movement of the encapsulated species are a fingerprint of the interaction between the fullerene cage and the metal complex. Based on the mode frequencies, the experimental value of the metal$-$cage bond strength can be determined and the amount of charge transfer from the cage to the core can be investigated \cite{lebedkin,jaffiol,krause}. It has been shown that some isolated fullerene cages, though electronically unstable, can bind strongly to the metal core and become stable \cite{qian}. Although certain gadolinium endofullerenes have been isolated and studied by X-ray diffraction, vibrational spectroscopy has not been performed on a number of species \cite{yang}.

In this paper, we present a detailed Raman and inelastic electron tunneling spectroscopy (IETS) analysis of Gd$_{3}$N@C$_{80}$, as well as conductance measurements demonstrating a Kondo effect. Raman analysis of Gd$_{3}$N@C$_{2n}$ (40 $\leq$ $n$ $\leq$ 44) has been performed, as well as Y$_{3}$N@C$_{80}$, Lu$_{3}$N@C$_{80}$, and Y$_{3}$N@C$_{88}$ for comparison. Vibrational assignments of the Gd$_{3}$N@C$_{80}$ molecule, I$_{h}-$C$^{6-}_{80}$ cage, and intrinsic Gd$_{3}$N cluster were carried out by density functional theory (DFT) and universal force field (UFF) calculations. Hindered rotations, shown by both theory and experiment, indicate the formation of a Gd$_{3}$N$-$C$_{80}$ bond which reduces the ideal icosahedral symmetry of the C$_{80}$ cage.

\section{Background}
An isolated Gd atom has a large magnetic moment of 8 $\mu_{B}$, where 7 $\mu_{B}$ is derived from localized $f$ electrons \cite{qian}. When embedded inside a fullerene, the Gd atom bonds through the delocalized $s$ and $d$ electrons and the localized $f$ states maintain their atomic spin moment \cite{qian}. The valence state of gadolinium has been determined experimentally by ESR (Electron Spin Resonance), XPS (X-ray Photoemission Spectroscopy) and X-ray diffraction to be Gd$^{3+}$ \cite{shinohara,shibata,mora,funasaka,laasonen,ding}. For the Gd$_{3}$N@C$_{2n}$ family, Gd can formally only be in the +1 charge state, because with three additional electrons, one from each Gd, the nitrogen is saturated in a formal $-3$ charge state. However, the highest charge state is Gd$^{3+}$ and therefore the whole Gd$_{3}$N cluster can donate six electrons to the cage resulting in (Gd$^{2+}$)$_{3}$N@C$^{6-}_{2n}$.

In the case of endohedral doping, the stability and symmetry of the cage depends on the properties of the encapsulated species, mostly on the amount of charge transfer. As a consequence, the isomer favored upon incorporation of an atom is usually different than the stable isomer of a hollow cage. Considering Gd$_{3}$N@C$_{80}$, the I$_{h}$ isomer of an isolated C$_{80}$ cage is unstable, having a four-fold degenerate HOMO occupied by only two electrons. The empty cage would undergo a Jahn-Teller distortion, leading to an occupied non-degenerate HOMO and a low-lying triply degenerate LUMO. However, the transfer of six additional electrons from the three Gd$^{3+}$ atoms stabilizes the I$_{h}-$C$_{80}$ by completely filling the HOMO orbital \cite{kobayashi2}. It is interesting to note that the intrinsic Gd$_{3}$N cluster and the C$_{80}$ cage do not exist independently, but when combined form a stable molecule with a closed shell structure.

The structure of Gd$_{3}$N@C$_{80}$ has been determined by X-ray diffraction and the Gd$_{3}$N unit is found to be pyramidal within the I$_{h}$ cage, with the three Gd atoms positioned over the centers of hexagons \cite{lu}. Unlike light mass pyramidal molecules, such as NH$_{3}$, the N atom is not able to tunnel through the Gd$_{3}$ plane due to a high energy barrier. However, when the intrinsic Gd$_{3}$N molecule is encapsulated in the C$_{80}$ cage the charge coupling effectively lowers the energy barrier to $\sim$ 91 meV, which allows tunneling to be possible \cite{qian}. As the cluster has a permanent electric dipole moment, the wagging mode involves a fluctuation in the overall dipole moment, which can be observed in the Raman spectra ($\sim$ 85 cm$^{-1}$) \cite{qian}. At room temperature, Gd$_{3}$N@C$_{80}$ is expected to be paramagnetic with the spin fluctuating between two different multiplicities (M = 0 and 22) \cite{lu}.

\section{Experiment and Results}
\subsection{Spectrophotometry and Raman Measurements}
UV-Vis spectrophotometry data for Gd$_{3}$N@C$_{2n}$ (40 $\leq$ $n$ $\leq$ 44) are shown in Fig. 1. The samples were isolated by HPLC and the purity was confirmed by mass spectrometry. The endofullerenes were suspended in toluene (99.999$\%$, Sigma-Aldrich\footnote[1]{Certain commercial equipment or materials are identified in this paper to foster understanding. Such identification does not imply recommendation or endorsement by the National Institute of Standards and Technology, nor does it imply that the materials or equipment identified are necessarily the best available for this purpose.}), which was background subtracted, and plotted as a function of wavelength.

\begin{figure}[h!]
\hskip 0.2cm\centerline{\epsfxsize=3.4in\epsfbox{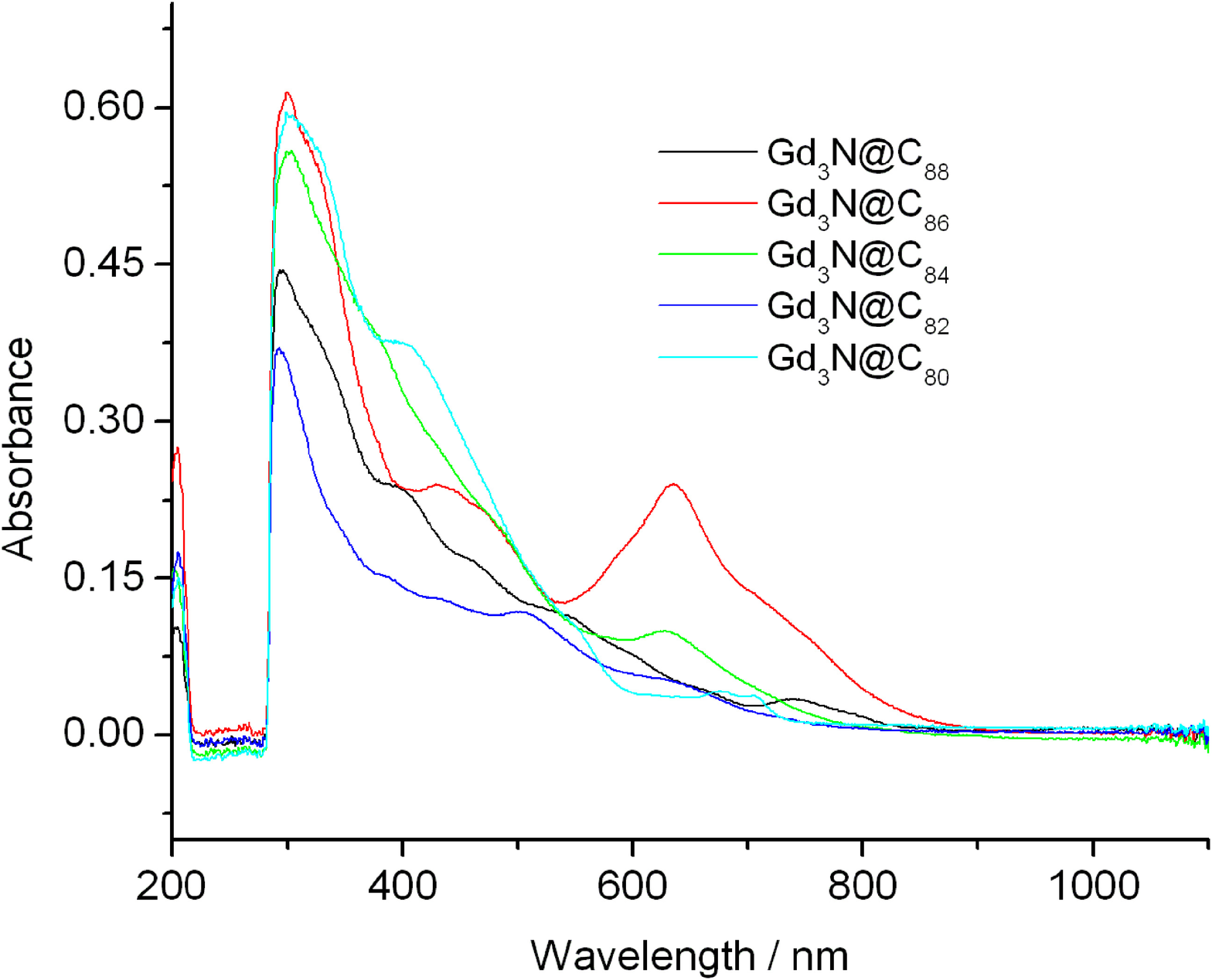}}
\vskip -0.4cm
\caption{Spectrophotometry data of Gd$_{3}$N@C$_{2n}$ (40 $\leq$ $n$ $\leq$ 44) in toluene. Unusually high absorption peak in Gd$_{3}$N@C$_{86}$ is observed.}
\label{fig1}
\end{figure}

The spectral onset for Gd$_{3}$N@C$_{80}$ is around 940 nm. The HOMO-LUMO transition has a doublet structure with absorption maxima at 680 and 710 nm. The strongest visible absorption of Gd$_{3}$N@C$_{80}$ is at 415 nm, and a shoulder is apparent at 555 nm \cite{krause2}. The spectral onset for Gd$_{3}$N@C$_{82}$ is around 950 nm and there are absorption peaks at 630 nm and 505 nm. The spectral onset for Gd$_{3}$N@C$_{84}$ is around 985 nm and there are absorption peaks at 915 nm and 635 nm. The spectral onset for Gd$_{3}$N@C$_{86}$ is around 1005 nm and there is an unusually strong absorption peak at 640 nm, as well as one at 430 nm. The spectral onset for Gd$_{3}$N@C$_{88}$ is around 960 nm and there are absorption peaks at 740 nm, 545 nm, 460 nm, and 405 nm.

For spectroscopic studies, 10 $\mu$g of Gd$_{3}$N@C$_{2n}$ (40 $\leq$ $n$ $\leq$ 44) were suspended in carbon disulfide (CS$_{2}$, 99.999$\%$, Sigma-Aldrich) and used to dropcoat a gold-covered silicon substrate. The resulting polycrystalline films were dried under ambient conditions and placed in a Linkam liquid nitrogen cryostat stage. Raman spectra were studied with 632.8 nm excitation from a HeNe laser. The scattered light was collected in the backscattering geometry by a triple-axis spectrometer (Jobin Yvon Horiba, T64000) equipped with a CCD detector. A resolution of 0.7 cm$^{-1}$ (1800 gr/mm) was used for all Raman measurements and all measurements used 2.5 mW/cm$^{2}$ laser power, 30 min. accumulation time, and were performed at 90 K.

The Raman spectra observed for Gd$_{3}$N@C$_{80}$ can be roughly divided into four parts: tangential C$_{80}$ modes are found between 1000 and 1600 cm$^{-1}$, a gap-like region from 815 to 1000 cm$^{-1}$, radial C$_{80}$ modes between 200 and 815 cm$^{-1}$, and a fourth group from 200 cm$^{-1}$ and below which has a counterpart only in the spectra of endohedral and polymeric fullerenes which exhibit low energetic Gd$-$cage, intermolecular, and center of mass (COM) modes. Low-energy Raman modes of Gd$_{3}$N@C$_{80}$ have been observed confirming the coupling between the core and cage by hindered rotation modes (Fig. 2).

\begin{figure}[h!]
\hskip 0.2cm\centerline{\epsfxsize=3.4in\epsfbox{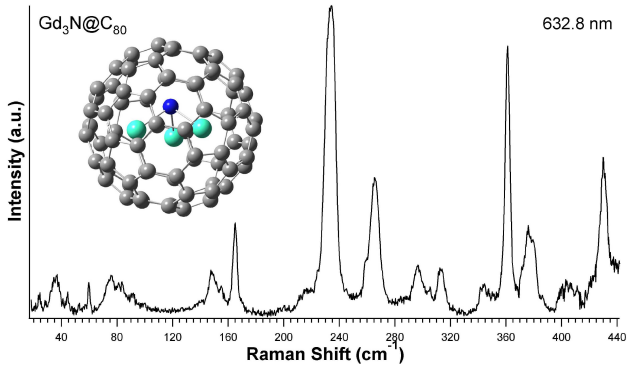}}
\vskip -0.4cm
\caption{Analysis of low-energy Raman lines of Gd$_{3}$N@C$_{80}$ taken at 90 K indicating a hindered rotation due to the coupling of the core complex to the cage. Inset gives the Gd$_{3}$N@C$_{80}$ structure.}
\label{fig2}
\end{figure}

Several prominent modes can be identified such as the C$_{80}$ H$_{g}$(1) squashing mode at 234 cm$^{-1}$, the H$_{g}$(2) twist mode at 361 cm$^{-1}$, and the A$_{g}$(1) breathing mode at 430 cm$^{-1}$. Below 200 cm$^{-1}$, several peaks are observed indicating a hindered rotation due to the bonding of Gd$_{3}$N to the C$_{80}$ cage. Other groups have seen evidence for the formation of a M$_{3}$N$-$C$_{80}$ bond which induces a significant reduction of the ideal I$_{h}$$-$C$_{80}$ symmetry \cite{krause3}. A peak at 83.6 cm$^{-1}$ appears to be the wagging mode of the nitrogen atom through the Gd$_{3}$ plane, predicted by theory \cite{qian}. Other COM modes are identified, where the cage$-$core complex translates relative to the other. The frequencies of these peaks were compared with the other gadolinium endofullerenes and are listed in Table I.

A similar procedure was performed to analyze the vibrational spectrum of the Gd$_{3}$N@C$_{2n}$ (41 $\leq$ $n$ $\leq$ 44) family. It can be clearly seen that the low-energy Raman modes are changed as the number of cage atoms increases (Fig. 3). A common mode for all samples arising from the coupling of the motion of the Gd complex and the fullerene cage seems to be apparent. The mode, defined from 140$-$165 cm$^{-1}$ and identified boldly in Table I, can be assigned to the Gd$-$cage stretching mode. This particular mode is similar to the vibrational mode reported for Gd@C$_{82}$, which exhibits a Raman frequency at 155 cm$^{-1}$ \cite{lebedkin}.

\begin{figure}[h!]
\hskip 0.2cm\centerline{\epsfxsize=3.4in\epsfbox{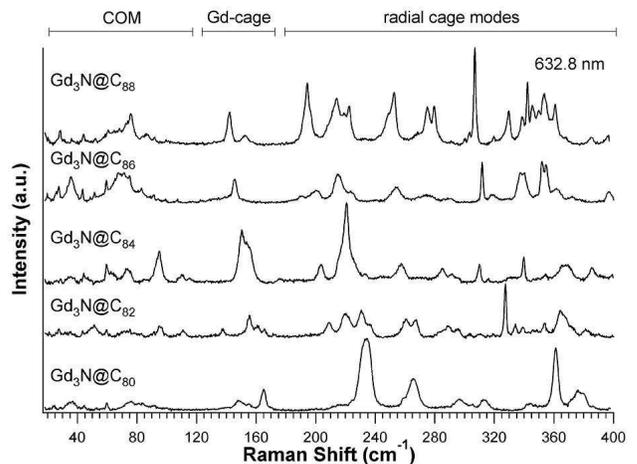}}
\vskip -0.4cm
\caption{{{Analysis of low-energy Raman lines of Gd$_{3}$N@C$_{2n}$ (40 $\leq$ $n$ $\leq$ 44) taken at 90 K indicating a hindered rotation due to the coupling of the core complex to the cage.}}}
\label{fig3}
\end{figure}

\begin{table}[ht]
\caption{Low-energy Raman frequencies (cm$^{-1}$) of Gd$_{3}$N@C$_{2n}$ (40 $\leq$ $n$ $\leq$ 44) below 200 cm$^{-1}$ at 90 K. Bold frequencies represent the Gd$-$cage stretching mode, which appears to be common for all gadolinium samples.}
\centering
\begin{tabular}{c c c c c}
\hline\hline
Gd$_{3}$N@C$_{80}$ & Gd$_{3}$N@C$_{82}$ & Gd$_{3}$N@C$_{84}$ & Gd$_{3}$N@C$_{86}$ & Gd$_{3}$N@C$_{88}$ \\
\hline\hline
$\bm{165.2}$ & 165.9 & 176.0 & $\bm{145.7}$ & 194.3 \\
148.4 & $\bm{155.4}$ & 156.2 & 106.9 & 152.8 \\
91.5 & 137.5 & $\bm{150.8}$ & 90.7 & $\bm{141.9}$ \\
83.6 & 111.2 & 115.5 & 83.2 & 91.9 \\
74.9 & 95.5 & 110.4 & 74.9 & 86.8 \\
59.8 & 59.4 & 95.1 & 59.4 & 76.2 \\
44.6 & 51.4 & 74.1 & 51.4 & 60.7 \\
35.8 & 27.4 & 60.2 & 43.5 & 51.9 \\
24.2 & & 44.6 & 35.8 & 43.9 \\
 & & 35.8 & 27.0 & 36.3 \\
 & & 28.6 & 19.8 & 28.3 \\
\hline
\end{tabular}
\label{tab1}
\end{table}

When plotting this mode against the square root of the reciprocal reduced mass, [1/$m_{reduced}$ (Gd$_{3}$N $-$ C$_{2n}$)]$^{1/2}$, a linear trend is found (Fig. 4). As the carbon cage mass increases, a linear frequency downshift is observed due to a decrease in the bond strength between the core complex and the cage. It is interesting to note that previous X-ray data has distinguished Gd$_{3}$N@C$_{80}$, which obeys the IPR and contains a pyramidal Gd$_{3}$N unit, from Gd$_{3}$N@C$_{82}$ and Gd$_{3}$N@C$_{84}$ which contain planar Gd$_{3}$ units and carbon cages that do not follow the IPR \cite{yang}. Although no X-ray or NMR data has been conducted for Gd$_{3}$N@C$_{86}$ and Gd$_{3}$N@C$_{88}$, the cages most likely follow the IPR and have D$_{3}$ and D$_{2}$ symmetry respectively \cite{zuo2,fu}. It is important to note that unlike Gd$_{3}$N@C$_{80}$, which has a preference for icosahedral symmetry (I$_{h}$), the other molecules have several IPR isomers that could be present in the Raman data. 

\begin{figure}[h!]
\hskip 0.2cm\centerline{\epsfxsize=3.4in\epsfbox{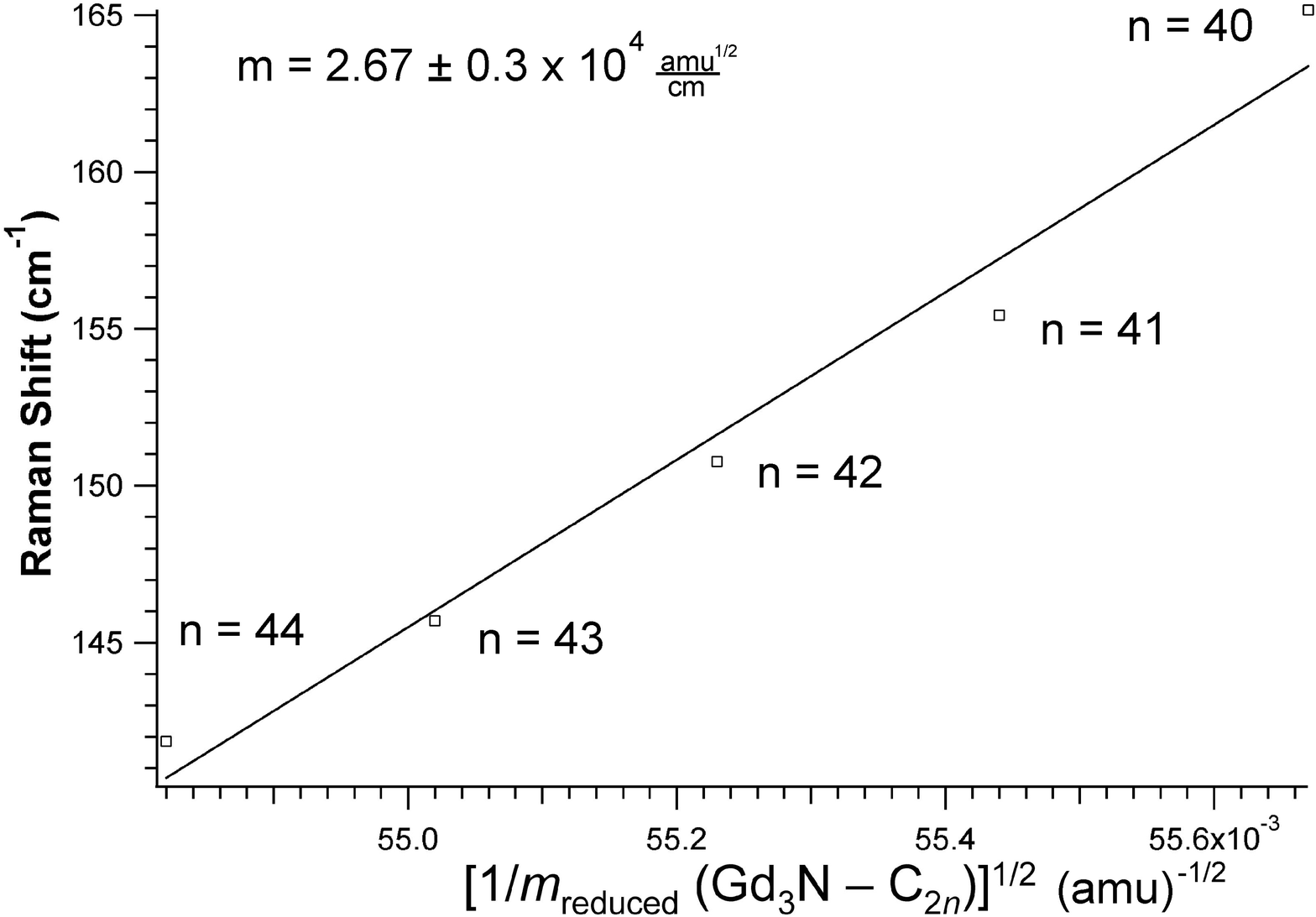}}
\vskip -0.4cm
\caption{Analysis of Gd$-$cage streching mode plotted against the square root of the reciprocal reduced mass. As the carbon mass increases, there is a linear downshift in the frequency.}
\label{fig4}
\end{figure}

Raman data of Y$_{3}$N@C$_{80}$ and Lu$_{3}$N@C$_{80}$ were analyzed in order to compare with the gadolinium complex Gd$_{3}$N@C$_{80}$ (Fig. 5). There is agreement above 220 cm$^{-1}$, which identifies the C$_{80}$ cage modes. The frequency of the M$-$cage stretching mode is increased to 195.6 cm$^{-1}$ for the yttrium complex, due to the change in reduced mass (32.7$\%$ decrease), and decreased to 155.9 cm$^{-1}$ for the lutetium complex, due to the change in reduced mass (7.0$\%$ increase). Assuming harmonic motion, a force constant k = 12.8 $\pm$ 0.37 N/cm can be derived, and it can be concluded that the frequency shift of the M$_{3}$N$-$C$_{80}$ oscillator is solely due to the change in metal mass \cite{krause4}. A splitting and separation of the M$-$cage stretching modes (123.4$-$195.6 cm$^{-1}$) is observed as the metal mass increases from yttrium to lutetium, as well as COM modes below 120 cm$^{-1}$.

\begin{figure}[h!]
\hskip 0.2cm\centerline{\epsfxsize=3.4in\epsfbox{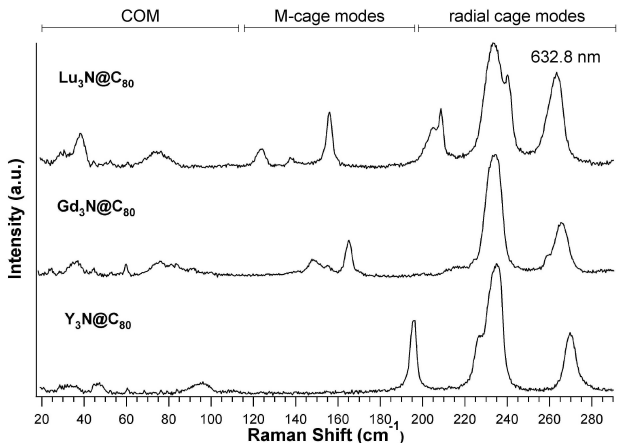}}
\vskip -0.4cm
\caption{Comparison of Y$_{3}$N@C$_{80}$, Gd$_{3}$N@C$_{80}$, and Lu$_{3}$N@C$_{80}$ low-energy Raman data taken at 90 K. Analysis of the data identifies C$_{80}$ cage modes, hindered rotation modes, and center of mass modes.}
\label{fig5}
\end{figure}

Raman data of Y$_{3}$N@C$_{88}$ has also been studied in order to compare with Gd$_{3}$N@C$_{88}$ (Fig. 6). There is excellent agreement above 250 cm$^{-1}$, despite several possible isomers, which identifies the C$_{88}$ cage modes. The frequency of the M$-$cage stretching mode is increased to 171.8 cm$^{-1}$ for the yttrium complex, due to the change in reduced mass (33.3$\%$ decrease). Assuming harmonic motion, a force constant k = 9.9 $\pm$ 0.12 N/cm can be derived and one can conclude that the frequency shift of the M$_{3}$N$-$C$_{88}$ oscillator is solely due to the change in metal mass \cite{krause4}. Several COM modes are observed in Gd$_{3}$N@C$_{88}$ at 93.5, 74.5, 60.2, 43.9, and 21.4 cm$^{-1}$ respectively.

\begin{figure}[h!]
\hskip 0.2cm\centerline{\epsfxsize=3.4in\epsfbox{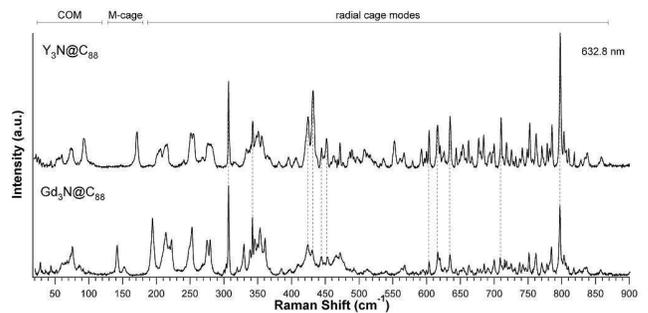}}
\vskip -0.4cm
\caption{Comparison of Y$_{3}$N@C$_{88}$ and Gd$_{3}$N@C$_{88}$ Raman data taken at 90 K. Analysis of the data identifies C$_{88}$ cage modes, hindered rotation modes and center of mass modes. Prominent peak correlations are denoted by dotted lines.}
\label{fig6}
\end{figure}

\subsection{IETS and Conductance Measurements}
The low temperature charge transport properties of Gd$_{3}$N@C$_{80}$ junctions were measured in a crossed-wire apparatus \cite{kushmerick}. A Gd$_{3}$N@C$_{80}$ layer is deposited from a $\sim$ 10 pM CS$_{2}$ solution onto a 10 $\mu$m diameter Au wire and mounted into a custom-built cryostat. A second 10 $\mu$m diameter Au wire serves as the top electrode. The cryostat assembly is evacuated, back filled with helium gas, and submerged into a liquid helium storage dewar. The junctions were formed after thermal equilibrium was reached and all transport measurements were performed at a temperature of 4.2 K. In Fig. 7, a zero-bias conductance peak commonly associated with Kondo scattering \cite{kondo} is clearly present. 

The Kondo effect is a many-body phenomenon that can arise from the coupling between a localized spin (Anderson impurity) and a sea of conduction electrons. It has been shown that molecules composed of magnetic ions can have a non-zero magnetic moment and give rise to Kondo scattering \cite{zhao,wahl,iancu}. The magnetic moment of the gadolinium ions is due to the localized 4$f$ states, which exhibit a non-vanishing magnetic moment and a localized single spin. The interaction of the $f$ electrons from the three Gd$^{2+}$ ions within the C$_{80}$ cage may in principle lead to the observed Kondo effect.

\begin{figure}[h!]
\hskip 0.2cm\centerline{\epsfxsize=3.4in\epsfbox{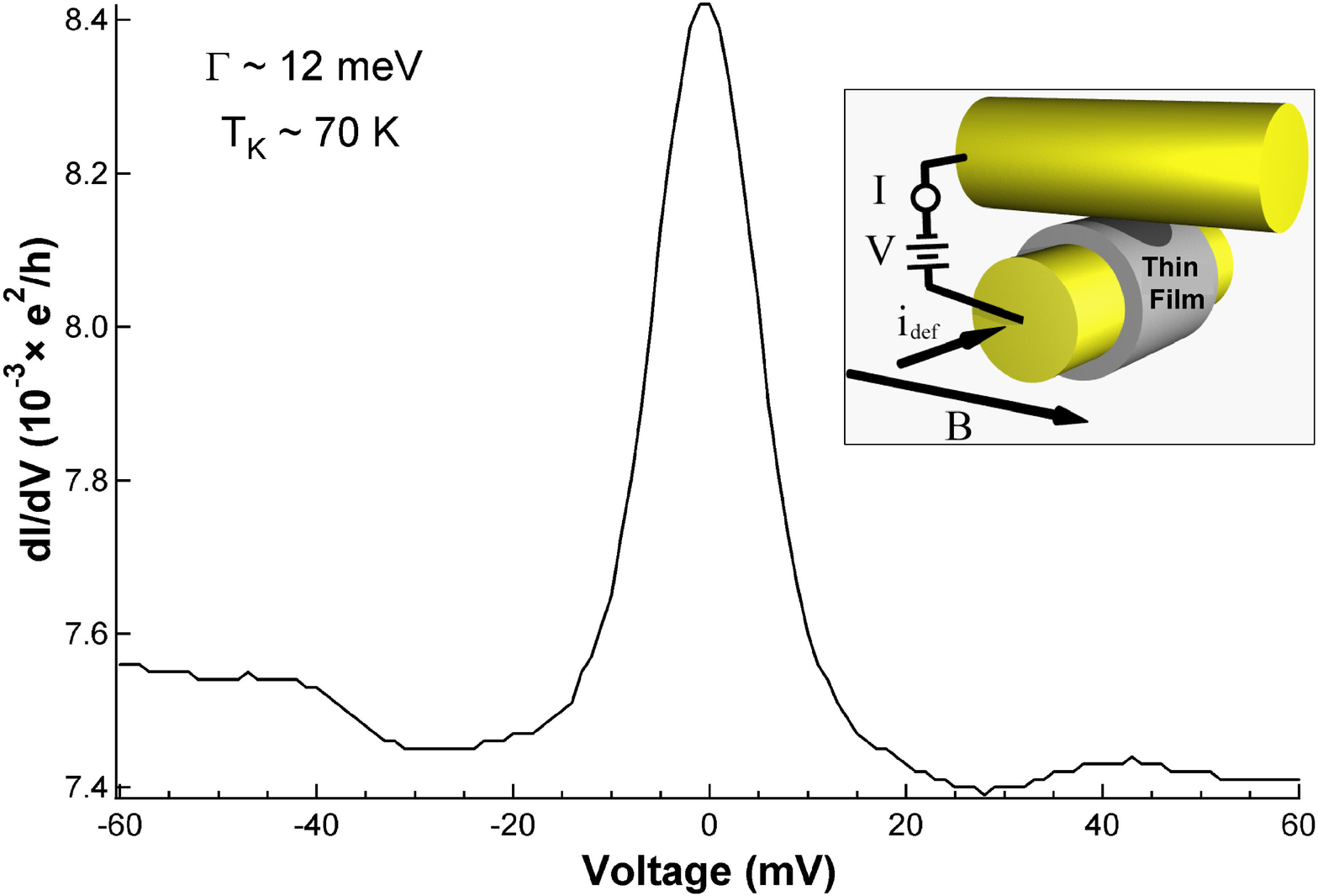}}
\vskip -0.4cm
\caption{{{Experimental conductance data of Kondo effect and zero-bias anomaly in Gd$_{3}$N@C$_{80}$ taken at 4.2 K. Inset adapted from \cite{kushmerick} shows the experimental setup: Au crossed-wire apparatus forms a junction with the Gd$_{3}$N@C$_{80}$ thin film.}}}
\label{fig7}
\end{figure}

It is important to note however, that similar Kondo scattering has been previously observed for junctions containing C$_{60}$ that do not have an intrinsic magnetic moment \cite{yu,parks}. Therefore, it is difficult to identify if the spin impurity is due to the C$_{80}$ cage, the gadolinium complex, or some combination of the two. Further investigation by magnetic STM (Scanning Tunneling Microscopy) or SPLEEM (Spin-polarized Low-energy Electron Microscopy) is required to have a more detailed analysis of the system.

Satellite peaks in the conductance data at $\pm$42 mV (339 cm$^{-1}$), similar to those reported \cite{parks} for C$_{60}$ at $\pm$33 mV (266 cm$^{-1}$), are also observed in the conductance spectra of the Gd$_{3}$N@C$_{80}$ junction. However, unlike previous groups who identified the peaks as the H$_{g}$(1) squashing mode for C$_{60}$, the observed satellite peaks cannot be definitively correlated with any intramolecular modes of Gd$_{3}$N@C$_{80}$.

\begin{figure}[h!]
\hskip 0.2cm\centerline{\epsfxsize=3.4in\epsfbox{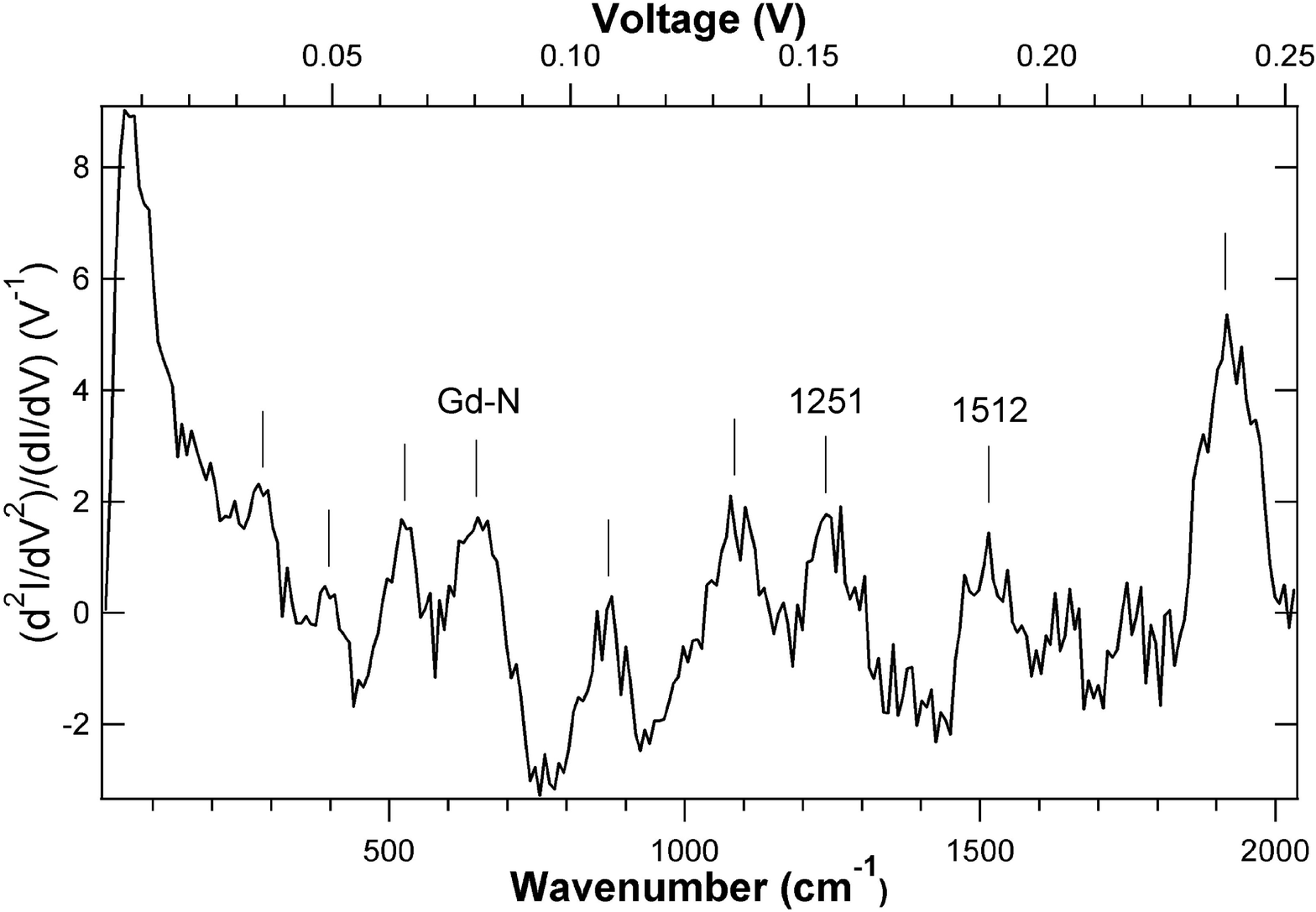}}
\vskip -0.4cm
\caption{{{Experimental IETS spectrum obtained for Gd$_{3}$N@C$_{80}$ with modulation amplitude V$_{\omega}$ = 4 mV, scan rate = 1.5 mV/s and time constant = 1 s. The anti-symmetric Gd$-$N stretch mode is identified at 81.6 mV (658 cm$^{-1}$) as well as Raman C$_{80}$ cage modes at 155.1 mV (1251 cm$^{-1}$) and 187.5 mV (1512 cm$^{-1}$).}}}
\label{fig8}
\end{figure}

Not all junctions formed exhibited the Kondo signature shown in Fig. 7. In fact, the transport properties were more or less evenly split between the Kondo resonance previously discussed and more conventional non-resonant tunneling and inelastic electron tunneling signatures \cite{jaklevic}. The inelastic electron tunneling spectrum of a Gd$_{3}$N@C$_{80}$ junction is shown in Fig. 8. There is a strong line at 81.6 mV (658 cm$^{-1}$) which has been identified previously by IR spectroscopy as the anti-symmetric Gd$-$N stretching vibration of the Gd$_{3}$N cluster \cite{krause2}. Raman C$_{80}$ cage modes are also observed at 155.1 mV (1251 cm$^{-1}$) and 187.5 mV (1512 cm$^{-1}$). The vibrational spectrum obtained by IETS is complimentary to Raman and IR measurements since although there are no strong selection rules \cite{hansma,beebe}, the molecular vibrations observed provides insight into the conduction pathways through the molecular species \cite{triosi}.

\section{Discussion}
In the case of C$_{80}$, its 12 pentagonal and 30 hexagonal faces can be arranged in seven different structures consistent with the IPR \cite{fowler}. The C$_{80}$ cage has elicited significant interest in part because after C$_{60}$, it has the smallest fullerene cage with an icosahedral isomer. The isolated C$_{80}$ molecule has 234 vibrational degrees of freedom: 80 $\times$ 3 = 240 degrees of freedom in total reduced by six corresponding to rotation and translation. Due to the icosahedral symmetry of the cage most of the vibrational modes are highly degenerate and fall into 62 distinct mode frequencies. The intramolecular modes can be analyzed and classified according to their symmetry \cite{dresselhaus} and from group theory it follows that 14 of the modes are Raman active (3A$_{g}$ and 11H$_{g}$), 6 are IR active (F$_{1u}$) and the remaining 42 are optically silent. Vibrational modes for the isolated I$_{h}$$-$C$^{6-}_{80}$ cage were calculated by hybrid DFT (B3LYP) utilizing the STO-3G/3-21G basis sets in the Gaussian 03 package \cite{frisch}. Pure C$_{80}$ cage vibrations should not be affected by the mass of the endohedral atoms \cite{krause3}. These values are compared with experimental and theoretical calculations (UFF) of Gd$_{3}$N@C$_{80}$ in Table II.

\begin{table}[ht]
\caption{Experimental and theoretical Raman frequencies (cm$^{-1}$) for Gd$_{3}$N@C$_{80}$, I$_{h}$$-$C$^{6-}_{80}$, and intrinsic Gd$_{3}$N.}
\centering
\begin{tabular}{c c c}
\hline\hline
Gd$_{3}$N@C$_{80}$ (Exp) & Gd$_{3}$N@C$_{80}$ (UFF) & I$_{h}$$-$C$^{6-}_{80}$ (DFT) \\
\hline\hline
430.4 A$_{g}$(1) & 428.5 A$_{g}$(1) & 429.7 A$_{g}$(1) \\
376.1 & 372.1 & 375.6 \\
361.3 H$_{g}$(2) & 370.7 H$_{g}$(2) & 355.3 H$_{g}$(2) \\
343.1 & 356.3 & 302.4 \\
313.2 & 302.2 & 298.1 \\
296.4 & 255.9 & 227.2 H$_{g}$(1) \\
265.7 & 229.1 & \\
234.5 H$_{g}$(1) & 217.9 H$_{g}$(1) & \\
165.2 Gd$-$cage & 141.8 Gd$-$cage & \\
148.4 & 116.3 & Gd$_{3}$N (UFF) \\
\cline{3-3}
91.5 & 115.4 & 747.8 \\
83.6 & 84.2 & 511.5 \\
74.9 & 79.2 & 112.1 \\
59.8 & 75.5 & 94.8 \\
44.6 & 46.7 & \\
35.8 & 44.3 & \\
24.2 & 25.4 & \\
\hline
\end{tabular}
\label{chp4tab2}
\end{table}

The C$^{6-}_{80}$ model accounts only for the charge transfer in endohedral complexes, which is a very simplified approximation of the core$-$cage interaction. In reality, the electronic properties of endofullerenes are determined additionally by the hybridization of the core$-$cage orbitals. Describing the presence of gadolinium only by adding electrons to the cage neglects some important effects arising from Gd$-$cage chemical bonding. In such a simple model the extra electrons are delocalized over the whole molecule, whereas more complex calculations for the endofullerenes predict localization of the charge in the vicinity of the chemisorption site.

The intrinsic Gd$_{3}$N molecule has six vibrational modes, two of which are degenerate. One mode with A$_{1}$ symmetry is a breathing mode or Gd$-$N symmetric stretch, where the Gd triangle expands and contracts. For the A$_{1}$ wagging mode or out-of-plane breathing mode, the Gd$_{3}$ moiety and the central nitrogen atom are displaced from the molecular plane in opposite directions. One of the E modes is a doubly-degenerate in-plane mode or asymmetric Gd$-$N stretching mode. The other E mode is an out-of-plane normal vibration or scissor mode, which involves changes in the Gd$-$N$-$Gd bond angle. All normal mode vibrations of the free Gd$_{3}$N molecule are both Raman and IR active.

The normal modes of the free Gd$_{3}$N molecule have been calculated by UFF. The doubly-degenerate scissor mode or $\delta$ (Gd$-$N$-$Gd) and asymmetric stretch or $\nu_{as}$ (Gd$-$N) have values at 94.8 cm$^{-1}$ and 747.8 cm$^{-1}$ respectively. The wagging mode or $\gamma$ (Gd$_{3}$$-$N) has the value of 511.5 cm$^{-1}$, whereas the breathing mode or $\nu_{s}$ (Gd$-$N) has a value of 112.1 cm$^{-1}$. The wagging mode is related to a Raman vibrational mode of GdN at 530 cm$^{-1}$ reported previously \cite{granville}.

\section{Conclusions}
We have described the structure and vibrational spectrum of the Gd$_{3}$N@C$_{2n}$ (40 $\leq$ $n$ $\leq$ 44) family. The endofullerenes were characterized by Raman spectroscopy, IETS, UV-Vis spectrophotometry, and molecular modeling (DFT and UFF). Tangential and radial cage modes, low-energy Gd$-$cage modes, intermolecular modes, and COM modes were identified in the Raman spectra. Hindered rotations were observed experimentally and calculated theoretically, indicating the formation of a Gd$_{3}$N$-$C$_{2n}$ (40 $\leq$ $n$ $\leq$ 44) bond which reduces the symmetry of the cage. As the carbon cage mass increases, a linear frequency downshift of the hindered rotation modes is observed due to a decrease in the bond strength between the core and the cage. Raman studies were conducted on Y$_{3}$N@C$_{80}$, Lu$_{3}$N@C$_{80}$ and Y$_{3}$N@C$_{88}$ for comparison with the Gd$_{3}$N@C$_{2n}$ family. Gd$_{3}$N@C$_{80}$ was further investigated by conductance measurements in a cross-wire apparatus at low temperature. A zero-bias conductance peak associated with Kondo scattering was observed. Based on these observations, we suggest that it is possible to utilize magnetic endofullerenes in magneto-electronic device applications, while characterizing the molecules through Raman and IETS. 

\section{Acknowledgements}
The authors thank Smitha Vasudevan for her assistance with molecular modeling. We are grateful for support of this work by the National Science Foundation [CHE-0443850 (H.C.D.), DMR-0507083 (H.C.D.)] and the National Institutes of Health [1R01-CA119371-01 (H.C.D.)]. A portion of this research at Oak Ridge National Laboratory's Center for Nanophase Materials Science was sponsored by the Scientific User Facilities Division, Office of Basic Energy Sciences, U.S. Department of Energy.

\end{document}